\documentclass{article}
\usepackage{amssymb}
\usepackage{amsfonts}
\usepackage{amsmath}

\input{tcilatex}
\begin{document}

\title{\textbf{Scalar-tensor theory with EGB term from Einstein Chern-Simons
gravity}}
\author{L. C\'{a}rdenas$^{1}$, V.C. Orozco$^{1}$, P. Salgado$^{2}$, D.
Salgado$^{1}$ and R. Salgado$^{1}$ \\
$^{1}$Departamento de F\'{\i}sica, Universidad de Concepci\'{o}n, \\
Casilla 160-C, Concepci\'{o}n, Chile\\
$^{2}$Instituto de Ciencias Exactas y Naturales, Facultad de Ciencias,\\
Univesidad Arturo Prat, Avda. Arturo Prat 2120, Iquique, Chile.}
\maketitle

\begin{abstract}
It is shown that the compactification \'{a} la Randall-Sundrum of the
so-called, five-dimensional Einstein-Chern-Simons action gravity leads to an
action for a four-dimensional scalar-tensor gravity that includes a
Gauss-Bonnet term, which belongs to a particular case of the action of the
Horndeski theory. The five-dimensional action includes new gravitational
degrees of freedom that were introduced requiring that the action be
invariant under symmetries greater than the usual Poincar\'{e} or (A)dS
symmetries, namely the so-called generalized Poincare algebras $\mathfrak{B}%
_{5}$.
\end{abstract}

\section{\textbf{Introduction}}

The construction of modified gravitational theories has been, in part,
motivated by the discovery of the accelerating expansion of the universe and
by the possibility that cosmic inflation might not be caused by a
cosmological constant but might require the introduction of additional
scalar degrees of freedom to the tensor degrees of freedom, capable of
significantly affecting gravitational dynamics. One way to achieve this is
by coupling the scalar field to the metric in various ways, which is
possible due to the simple behavior of the scalar field under Lorentz
transformations. This procedure gave rise to scalar-tensor theories that
correspond to generalizations of Jordan-Brans-Dicke theories and to theories
that lead to second-order field equations known as Horndeski theories \cite%
{horn} (see also \cite{galileo}).

In reference \cite{garciabellido}, the space of the scalar-tensor theories
was studied, and making use of the language of differential forms, it was
found that said space has a finite and closed base of Lagrangians, which
allow classifying the scalar-tensor theories up to now known.

On the other hand, the inclusion of new gravitational degrees of freedom can
be done via the construction of gravitational theories, invariant under
symmetries greater than the usual Poincar\'{e} or (A)dS symmetries, namely
the so-called generalized Poincare algebras \cite{saka}-\cite{quince}\textbf{%
.}

The main purpose of this letter is to show that the compactification, \'{a}
la Randall-Sundrum, \cite{rand1},\cite{rand2}, of the so-called,
five-dimensional Einstein-Chern-Simons action gravity leads to an action for
a four-dimensional scalar-tensor theory of gravity that includes a
Gauss-Bonnet term, which belongs to a particular case of the action
corresponding to the Lagrangian ($48$) of Ref.\textbf{\ }\cite{garciabellido}%
. It should be noted that this same compactification procedure applied to
the five-dimensional AdS-Chern-Simons action \cite{cham1}-\cite{zan2}, leads
to the Einstein-Hilbert-Cartan action in four dimensions \cite{lepe} (see
appendix $B$).

The Einstein-Chern-Simons gravity \cite{tres} is a gauge theory whose
Lagrangian density is given by a 5-dimensional Chern-Simons form for the so
called $\mathfrak{B}_{5}$ algebra. This algebra can be obtained from the
Anti-de Sitter algebra and a particular semigroup $S$ by means of the $S$%
-expansion procedure introduced in Refs. \cite{saka}-\cite{tres}.\ The
content fields induced by the $\mathfrak{B}_{5}$ algebra includes the
vielbein $e^{a}$, the spin connection $\omega ^{ab}$, and two extra bosonic
fields $h^{a}$ and $k^{ab}.$ The $5$-dimensional Chern-Simons Lagrangian for
the $\mathcal{B}_{5}$ algebra is given by \cite{tres}%
\begin{eqnarray}
L_{ChS}^{(5)}[e,\omega ,h,k] &=&\alpha _{1}l^{2}\varepsilon
_{abcde}R^{ab}R^{cd}e^{e}  \notag \\
&&+\alpha _{3}\varepsilon _{abcde}\left( \frac{2}{3}%
R^{ab}e^{c}e^{d}e^{e}+2l^{2}k^{ab}R^{cd}T^{\text{ }e}+l^{2}R^{ab}R^{cd}h^{e}%
\right)  \notag \\
&&+d\hat{B}_{EChS}^{(4)},  \label{1t}
\end{eqnarray}%
where $\alpha _{1}$, $\alpha _{3}$ are parameters of the theory, $R^{ab}=%
\mathrm{d}\omega ^{ab}+\omega _{\text{ }c}^{a}\omega ^{cb}$ and $T^{a}=%
\mathrm{d}e^{a}+\omega _{\text{ }c}^{a}e^{c}$, leads to the standard
five-dimensional general relativity without cosmological constant in the
limit where the coupling constant $l$ tends to zero while keeping the
Newton's constant fixed. \ It should be noted that the kinetic terms for the 
$h^{a}$ and $k^{ab}$ fields are present only in the surface term of the
Lagrangian $\hat{B}_{EChS}^{(4)}$ shown in (\ref{1t}).

The Letter is organized as follows: In Section $2$, it is shown that the
Randall-Sundrum compactification of the Einstein-Chern-Simons action in five
dimensions leads to an action for a four-dimensional scalar-tensor theory of
gravity that includes a Gauss-Bonnet term, which belongs to the particular
case of the action corresponding to the Lagrangian ($48$) of Ref.\textbf{\ }%
\cite{garciabellido}\textbf{\ }(see appendix $C$). In Section $3$ we will
study the action found in Section $2$ for the case of a maximally symmetric
space-time. A discussion about the results are presented in section $4$.
Three appendices to make this article a self-contained work are also
considered.

\section{\textbf{Scalar-tensor gravity with Gauss-Bonnet term from
Einstein-Chern-Simons gravity}}

Taking into account that $T^{a}=\mathrm{D}e^{a}$ the Lagrangian (\ref{1t})
can be written in the form

\begin{eqnarray}
L_{ChS}^{(5)}[e,\omega ,h,k] &=&\alpha _{1}\ell ^{2}\epsilon
_{abcde}R^{ab}R^{cd}e^{e}+\alpha _{3}\epsilon _{abcde}\left( \frac{2}{3}%
R^{ab}e^{c}e^{d}e^{e}\right.  \notag \\
&&\left. +2\ell ^{2}Dk^{ab}R^{cd}e^{e}+\ell ^{2}R^{ab}R^{cd}h^{e}\right) .
\label{ecu1}
\end{eqnarray}

From equation (\ref{ecu1}) we can see that the Lagrangian contains four
terms that we will denote as $L_{1},L_{2},L_{3},L_{4},$ where $L_{1}$
corresponds to the Gauss-Bonnet term and $L_{2}$ corresponds to the
Einstein-Hilbert term. Following Ref. \cite{rd,gomez} we replace the results
(\ref{s2}) and (\ref{105t}), shown in the Appendix $A$, in (\ref{ecu1}), we
find

\begin{eqnarray}
L_{1} &=&\alpha _{1}\ell ^{2}\epsilon _{abcde}R^{ab}R^{cd}e^{e}  \notag \\
&&\alpha _{1}\ell ^{2}r_{c}d\phi \left[ \tilde{\epsilon}_{mnpq}\tilde{R}^{mn}%
\tilde{R}^{pq}-2\frac{e^{2f(\phi )}}{r_{c}^{2}}\left( 2f^{\prime \prime
}+3f^{\prime 2}\right) \tilde{\epsilon}_{mnpq}\tilde{R}^{mn}\tilde{e}^{p}%
\tilde{e}^{q}+\right.  \notag \\
&&\left. \frac{e^{4f(\phi )}}{r_{c}^{4}}\left( 4f^{\prime \prime
}+5f^{\prime 2}\right) f^{\prime 2}\tilde{\epsilon}_{mnpq}\tilde{e}^{m}%
\tilde{e}^{n}\tilde{e}^{p}\tilde{e}^{q}\right] ,  \label{ecu2}
\end{eqnarray}

\begin{eqnarray}
L_{2} &=&\frac{2\alpha _{3}}{3}\epsilon _{abcde}R^{ab}e^{c}e^{d}e^{e}  \notag
\\
&=&\frac{2}{3}\alpha _{3}r_{c}d\phi \left[ 3e^{2f(\phi )}\tilde{\epsilon}%
_{mnpq}\tilde{R}^{mn}\tilde{e}^{p}\tilde{e}^{q}\right.  \notag \\
&&\left. -\frac{e^{4f(\phi )}}{r_{c}^{2}}\left( 2f^{\prime \prime
}+5f^{\prime 2}\right) \tilde{\epsilon}_{mnpq}\tilde{e}^{m}\tilde{e}^{n}%
\tilde{e}^{p}\tilde{e}^{q}\right] ,  \label{ecu3}
\end{eqnarray}

\begin{eqnarray}
L_{3} &=&2\alpha _{3}\ell ^{2}\epsilon _{abcde}\left( Dk^{ab}\right)
R^{cd}e^{e}  \notag \\
&=&-2\alpha _{3}\ell ^{2}\frac{e^{2f(\phi )}}{r_{c}^{2}}\left( 2f^{\prime
\prime }+3f^{\prime 2}\right) r_{c}d\phi \tilde{\epsilon}_{mnpq}\left( D%
\tilde{k}^{mn}\right) \tilde{e}^{p}\tilde{e}^{q}  \notag \\
&&+2\alpha _{3}\ell ^{2}r_{c}d\phi \tilde{\epsilon}_{mnpq}\tilde{R}%
^{mn}\left( D\tilde{k}^{pq}\right) ,  \label{ecu11}
\end{eqnarray}

\begin{eqnarray}
L_{4} &=&\alpha _{3}\ell ^{2}\epsilon _{abcde}R^{ab}R^{cd}h^{e}  \notag \\
&=&-4\alpha _{3}\ell ^{2}\left( f^{\prime \prime }+f^{\prime 2}\right) \frac{%
e^{f(\phi )}}{r_{c}^{2}}e^{g(\phi )}r_{c}d\phi \tilde{\epsilon}_{mnpq}\tilde{%
R}^{mn}\tilde{e}^{p}\tilde{h}^{q}  \notag \\
&&+4\alpha _{3}\ell ^{2}\left( f^{\prime \prime }+f^{\prime 2}\right) \frac{%
e^{3f(\phi )}}{r_{c}^{4}}f^{\prime 2}e^{g(\phi )}r_{c}d\phi \tilde{\epsilon}%
_{mnpq}\tilde{e}^{m}\tilde{e}^{n}\tilde{e}^{p}\tilde{h}^{q},  \label{ecu17}
\end{eqnarray}%
where we have taken into account that $k^{m4}=0$ and we have used\newline

\begin{equation}
h^{m}\left( \phi ,\tilde{x}\right) =e^{g(\phi )}\tilde{h}^{m}(\tilde{x}%
),\quad h^{4}=0.  \label{ecu18}
\end{equation}

Replacing (\ref{ecu2},\ref{ecu3},\ref{ecu11},\ref{ecu17}) in (\ref{ecu1})
and integrating over the fifth dimension we find

\begin{eqnarray}
\tilde{S}^{4D}\left[ \tilde{e},\tilde{k},\tilde{h}\right]  &=&\int_{\Sigma
_{4}}\tilde{\epsilon}_{mnpq}\left\{ \alpha \tilde{R}^{mn}\tilde{e}^{p}\tilde{%
e}^{q}+\beta \tilde{e}^{m}\tilde{e}^{n}\tilde{e}^{p}\tilde{e}^{q}+\gamma
\left( D\tilde{k}^{mn}\right) \tilde{e}^{p}\tilde{e}^{q}\right.   \notag \\
&&\left. +\tau \tilde{R}^{mn}\left( D\tilde{k}^{pq}\right) +\chi \tilde{R}%
^{mn}\tilde{e}^{p}\tilde{h}^{q}+\xi \tilde{e}^{m}\tilde{e}^{n}\tilde{e}^{p}%
\tilde{h}^{q}\right\} ,  \label{ecu19}
\end{eqnarray}%
where\newline
\begin{eqnarray}
\alpha  &=&\frac{2\alpha _{1}\ell ^{2}\pi }{r_{c}}+2\pi \alpha _{3}r_{c},%
\text{ \ \ \ \ \ }\beta =-\frac{11\alpha _{1}\ell ^{2}\pi }{4r_{c}^{3}}+%
\frac{\alpha _{3}\ell ^{2}\pi }{2r_{c}},  \notag \\
\gamma  &=&\frac{2\alpha _{3}\ell ^{2}\pi }{r_{c}},\text{ \ \ \ }\tau =4\pi
\alpha _{3}\ell ^{2}r_{c},\text{ \ \ }\chi =\frac{4\alpha _{3}\ell ^{2}\pi }{%
r_{c}},\text{ \ \ \ \ }\xi =-\frac{\alpha _{3}\ell ^{2}\pi }{r_{c}^{3}},
\label{ecu20}
\end{eqnarray}%
are coefficients that were calculated by choosing $f(\phi )=g(\phi )=\ln
(\sin \phi )$. Note that when $l\longrightarrow 0$ the action results in the
Einstein-Hilbert action in $4D$.

The tensor form of the action (\ref{ecu19}) can be obtained by remembering
that the first two terms can be written as

\begin{equation}
\tilde{\epsilon}_{mnpq}\tilde{R}^{mn}\tilde{e}^{p}\tilde{e}^{q}=4\sqrt{-%
\tilde{g}}\tilde{R}d^{4}\tilde{x},  \label{1}
\end{equation}

\begin{equation}
\tilde{\epsilon}_{mnpq}\tilde{e}^{m}\tilde{e}^{n}\tilde{e}^{p}\tilde{e}%
^{q}=4!\sqrt{-\tilde{g}}d^{4}\tilde{x},  \label{2}
\end{equation}%
where $\tilde{g}$ is the determinant of the $4$-dimensional metric $\tilde{g}%
_{\mu \nu }$ and $\tilde{R}$ is the $4$-dimensional Ricci scalar. The
remaining terms are obtained after some calculations. The results are given
by

\begin{equation}
\tilde{\epsilon}_{mnpq}\left( D\tilde{k}^{mn}\right) \tilde{e}^{p}\tilde{e}%
^{q}=4\sqrt{-\tilde{g}}D_{i}\tilde{k}_{\ \ j}^{ij}d^{4}\tilde{x},  \label{3}
\end{equation}

\begin{eqnarray}
\tilde{\epsilon}_{mnpq}\tilde{R}^{mn}D\tilde{k}^{pq} &=&4\sqrt{-\tilde{g}}%
\left( \tilde{R}_{\ \ \rho \sigma }^{\mu \nu }D_{\mu }\tilde{k}_{\ \ \nu
}^{\rho \sigma }-2\tilde{R}_{\ \nu }^{\mu }D_{\rho }\tilde{k}_{\ \ \mu
}^{\rho \nu }\right.   \notag \\
&&\left. -2\tilde{R}_{\ \nu }^{\mu }D_{\mu }\tilde{k}_{\ \ \rho }^{\nu \rho
}+\tilde{R}D_{\mu }\tilde{k}_{\ \ \nu }^{\mu \nu }\right) d^{4}\tilde{x},
\label{4}
\end{eqnarray}

\begin{equation}
\tilde{\epsilon}_{mnpq}\tilde{R}^{mn}\tilde{e}^{p}\tilde{h}^{q}=2\sqrt{-%
\tilde{g}}\left( \tilde{R}\tilde{h}-2\tilde{R}_{\ \nu }^{\mu }\tilde{h}_{\
\mu }^{\nu }\right) d^{4}\tilde{x},  \label{5}
\end{equation}

\begin{equation}
\tilde{\epsilon}_{mnpq}\tilde{e}^{m}\tilde{e}^{n}\tilde{e}^{p}\tilde{h}^{q}=6%
\sqrt{-\tilde{g}}\tilde{h}d^{4}\tilde{x},  \label{6}
\end{equation}%
where 
\begin{equation}
\tilde{h}^{m}=\tilde{h}_{\ \mu }^{m}d\tilde{x}^{\mu },\text{ \ }\tilde{e}%
^{m}=\tilde{e}_{\ \mu }^{m}d\tilde{x}^{\mu },\text{ \ \ }\tilde{\omega}^{mn}=%
\tilde{\omega}_{\ \mu }^{mn}d\tilde{x}^{\mu },\text{ \ }\tilde{k}^{mn}=%
\tilde{k}_{\ \mu }^{mn}d\tilde{x}^{\mu }.\text{\ \ }  \label{7}
\end{equation}

Introducing (\ref{1}-\ref{6}) in the action (\ref{ecu19}) we have that this
one, in tensor language, takes the form

\begin{eqnarray}
\tilde{S} &=&\int_{\Sigma _{4}}\sqrt{-\tilde{g}}\left[ 4\alpha \tilde{R}%
+24\beta +4\gamma D_{\mu }\tilde{k}_{\ \ \nu }^{\mu \nu }\right.   \notag \\
&&+4\tau \left( \tilde{R}_{\ \ \rho \sigma }^{\mu \nu }D_{\mu }\tilde{k}_{\
\ \nu }^{\rho \sigma }-2\tilde{R}_{\ \nu }^{\mu }D_{\rho }\tilde{k}_{\ \ \mu
}^{\rho \nu }-2\tilde{R}_{\ \nu }^{\mu }D_{\mu }\tilde{k}_{\ \ \rho }^{\nu
\rho }+\tilde{R}D_{\mu }\tilde{k}_{\ \ \nu }^{\mu \nu }\right)   \notag \\
&&\left. +2\chi \left( \tilde{R}\tilde{h}-2\tilde{R}_{\ \nu }^{\mu }\tilde{h}%
_{\ \mu }^{\nu }\right) +6\xi \tilde{h}\right] d^{4}\tilde{x}.  \label{ecu66}
\end{eqnarray}

\section{\textbf{Action for a maximally symmetric space-time}}

If we consider a maximally symmetric space-time (for instance, the de-Sitter
space), the equation $13.4.6$ of Ref. \cite{weinberg} allows us to write%
\begin{equation}
\tilde{h}^{q}=F\left( \varphi \right) \tilde{e}^{q},\text{ \ }\tilde{k}%
^{mn}=G\left( \varphi \right) \tilde{\omega}^{mn},  \label{ecu21}
\end{equation}%
so that%
\begin{eqnarray}
\tilde{\epsilon}_{mnpq}\tilde{R}^{mn}\tilde{e}^{p}\tilde{h}^{q} &=&F\left(
\varphi \right) \tilde{\epsilon}_{mnpq}\tilde{R}^{mn}\tilde{e}^{p}\tilde{e}%
^{q},  \label{ecu22} \\
\tilde{\epsilon}_{mnpq}\tilde{e}^{m}\tilde{e}^{n}\tilde{e}^{p}\tilde{h}^{q}
&=&F\left( \varphi \right) \tilde{\epsilon}_{mnpq}\tilde{e}^{m}\tilde{e}^{n}%
\tilde{e}^{p}\tilde{e}^{q},  \label{ecu23} \\
\tilde{\epsilon}_{mnpq}D\tilde{k}^{mn}\tilde{e}^{p}\tilde{e}^{q} &=&\frac{DG%
}{G}\tilde{\epsilon}_{mnpq}\tilde{k}^{mn}\tilde{e}^{p}\tilde{e}^{q}+G\tilde{%
\epsilon}_{mnpq}\tilde{R}^{mn}\tilde{e}^{p}\tilde{e}^{q},  \label{ecu24} \\
\tilde{\epsilon}_{mnpq}\tilde{R}^{mn}D\tilde{k}^{pq} &=&\frac{DG}{G}\tilde{%
\epsilon}_{mnpq}\tilde{R}^{mn}\tilde{k}^{pq}+G\tilde{\epsilon}_{mnpq}\tilde{R%
}^{mn}\tilde{R}^{pq}.  \label{ecu25}
\end{eqnarray}%
Using (\ref{ecu22}-\ref{ecu25}) we find that the action (\ref{ecu19}) takes
the fom 
\begin{eqnarray}
\tilde{S}\left[ \tilde{e},\tilde{k},\tilde{h}\right] &=&\int_{\Sigma _{4}}%
\tilde{\epsilon}_{mnpq}\left\{ V_{1}\left( \varphi \right) \tilde{R}^{mn}%
\tilde{e}^{p}\tilde{e}^{q}+V_{2}\left( \varphi \right) \tilde{e}^{m}\tilde{e}%
^{n}\tilde{e}^{p}\tilde{e}^{q}\right.  \notag \\
&&\left. +V_{3}\left( \varphi \right) \tilde{k}^{mn}\tilde{e}^{p}\tilde{e}%
^{q}+V_{4}\left( \varphi \right) \tilde{R}^{mn}\tilde{k}^{pq}+V_{5}\left(
\varphi \right) \tilde{R}^{mn}\tilde{R}^{pq}\right\} ,  \label{ecu26}
\end{eqnarray}%
where $V_{1}\left( \varphi \right) =\alpha +\gamma G\left( \varphi \right)
+\chi F\left( \varphi \right) ,$ \ $V_{2}\left( \varphi \right) =\beta +\xi
F\left( \varphi \right) ,$ \ $V_{3}\left( \varphi \right) =\gamma DG/G,$ $%
V_{4}\left( \varphi \right) =\tau DG/G,$ $V_{5}\left( \varphi \right) =\tau
G.$

The tensor form of the action (\ref{ecu26}) can be obtained, taking into
account, in addition to (\ref{1}) and (\ref{2}), that

\begin{eqnarray}
V_{3}\left( \varphi \right) \tilde{\epsilon}_{mnpq}\tilde{k}^{mn}\tilde{e}%
^{p}\tilde{e}^{q} &=&\gamma \frac{DG}{G}\tilde{\epsilon}_{mnpq}\tilde{k}^{mn}%
\tilde{e}^{p}\tilde{e}^{q}=\gamma 4\sqrt{-\tilde{g}}\frac{D_{\mu }G}{G}%
\left\{ \tilde{k}_{\ \ \text{\ }\nu }^{\mu \nu }\right\} d^{4}\tilde{x}, \\
V_{4}\left( \varphi \right) \tilde{\epsilon}_{mnpq}\tilde{R}^{mn}\tilde{k}%
^{pq} &=&\tau \frac{DG}{G}\tilde{R}^{mn}\tilde{k}^{pq}=\tau 4\sqrt{-\tilde{g}%
}\frac{D_{\mu }G}{G}\left\{ \tilde{R}\tilde{k}_{\ \ \sigma }^{\mu \sigma }+2%
\tilde{R}_{\ \text{\ }\nu }^{\sigma }\tilde{k}_{\ \ \text{\ }\sigma }^{\nu
\mu }\right.   \notag \\
&&\left. +\tilde{R}_{\ \ \rho \sigma }^{\mu \nu }\tilde{k}_{\ \ \nu }^{\rho
\sigma }-2\tilde{R}_{\ \nu }^{\mu }\tilde{k}_{\ \ \sigma }^{\nu \sigma
}\right\} d^{4}\tilde{x}  \notag \\
&=&-\tau 4\sqrt{-\tilde{g}}\frac{\left( D_{\mu }G\right) }{G}\left[ 2\tilde{G%
}^{\mu \nu }\tilde{k}_{\nu \ \sigma }^{\ \sigma }-2\tilde{R}_{\ \nu
}^{\sigma }\tilde{k}_{\ \ \sigma }^{\nu \mu }-\tilde{R}_{\ \ \nu \rho }^{\mu
\sigma }\tilde{k}_{\ \ \sigma }^{\nu \rho }\right] d^{4}\tilde{x},  \notag \\
&& \\
V_{5}\left( \varphi \right) \tilde{\epsilon}_{mnpq}\tilde{R}^{mn}\tilde{R}%
^{pq} &=&\tau G\sqrt{-g}\left( R^{2}-4R_{\mu \nu }R^{\mu \nu }+R_{\mu \nu
\rho \sigma }R^{\mu \nu \rho \sigma }\right) d^{4}x.  \label{ecu27}
\end{eqnarray}%
In fact, introducing (\ref{1}), (\ref{2}) and (\ref{ecu27}) in \ref{ecu66},
we obtain%
\begin{eqnarray}
\tilde{S}\left[ \tilde{e},\tilde{k},\tilde{h}\right]  &=&\int_{\Sigma
_{4}}d^{4}\tilde{x}\sqrt{-\tilde{g}}\left[ \tau G\left( R^{2}-4R_{\mu \nu
}R^{\mu \nu }+R_{\mu \nu \rho \sigma }R^{\mu \nu \rho \sigma }\right)
\right.   \notag \\
&&+4\left( \alpha +\gamma G+\chi F+\tau \frac{D_{\mu }G}{G}\tilde{k}_{\ \
\sigma }^{\mu \sigma }\right) \tilde{R}  \notag \\
&&+\frac{8\tau }{G}\left( D_{\sigma }G\tilde{k}_{\ \ \text{\ }\mu }^{\nu
\sigma }-D_{\mu }G\tilde{k}_{\ \ \sigma }^{\nu \sigma }\right) \tilde{R}_{\
\nu }^{\mu }+\frac{4\tau }{G}D_{\mu }G\tilde{k}_{\ \ \nu }^{\rho \sigma }%
\tilde{R}_{\ \ \rho \sigma }^{\mu \nu }  \notag \\
&&\left. +24\left( \beta +\xi F+\gamma \frac{D_{\mu }G}{6G}\tilde{k}_{\ \ 
\text{\ }\nu }^{\mu \nu }\right) \right] ,  \notag \\
&&  \label{ecu28}
\end{eqnarray}%
where $G=G(\varphi )$, $F=F\left( \varphi \right) $ and $\alpha ,\beta
,\gamma ,\xi ,\chi ,\tau $ are arbitrary constants.

Comparing (\ref{ecu28}) with (\ref{a4}), we see that the coefficients of the
Gauss-Bonnet term $\tau G\left( \phi \right) $, of the Ricci scalar $4\left(
\alpha +\gamma G+\chi F+\tau \frac{D_{\mu }G}{G}\tilde{k}_{\ \ \sigma }^{\mu
\sigma }\right) $, of the Ricci tensor $\left( 8\tau /G\right) \left(
D_{\sigma }G\tilde{k}_{\ \ \text{\ }\mu }^{\nu \sigma }-D_{\mu }G\tilde{k}%
_{\ \ \sigma }^{\nu \sigma }\right) $, of the Riemann tensor $\left( 4\tau
/G\right) D_{\mu }G\tilde{k}_{\ \ \nu }^{\rho \sigma }$, could be
identified, respectively, with the coefficients $E_{6},$ $2E_{6,X}\left( %
\left[ \Phi \right] ^{2}-\left[ \Phi ^{2}\right] \right) ,$ $8E_{6,X}\left( %
\left[ \Phi \right] \Phi _{ab}-\Phi _{ab}^{2}\right) ,$ $4E_{6,X}\Phi
_{ab}\Phi _{cd}$, $8E_{6,X}\left( \left[ \Phi \right] \Phi _{ab}-\Phi
_{ab}^{2}\right) ,$ $4E_{6,X}\Phi _{ab}\Phi _{cd}$. Finally, the coefficient 
$24\left( \beta +\xi F+\gamma \left( D_{\mu }G/6G\right) \tilde{k}_{\ \ 
\text{\ }\nu }^{\mu \nu }\right) $ could be identified with $%
(1/3)E_{6,XX}\left( \left[ \Phi \right] ^{4}-6\left[ \Phi \right] ^{2}\left[
\Phi ^{2}\right] +3\left[ \Phi ^{2}\right] ^{2}+8\left[ \Phi \right] \left[
\Phi ^{3}\right] -6\left[ \Phi ^{4}\right] \right) $.

This correspondence leads in principle to a probable interpretation of the
fields $\tilde{k}_{\ \ \sigma }^{\mu \sigma }$ and $\tilde{k}_{\ \ \nu
}^{\rho \sigma }$ as first derivative and third derivative of a scalar field
respectively.

\section{\textbf{Concluding Remarks }}

In this work it was found that the introduction of new degrees of freedom in
the action of AdS-Chern-Simons gravity via the expansion procedure \cite%
{uno,dos,tres}, leads, after a compactification, \`{a} la Randall-Sundrum,
to an action for a four-dimensional scalar-tensor gravity that includes a
Gauss-Bonnet term, which belongs to a family of actions of the Horndeski
theory.

Indeed, the four-dimensional scalar-tensor theory of gravity was obtained by
compactification of the five-dimensional Einstein-Chern-Simons action
gravity. It corresponds to a particular case of the Lagrangian ($48$) of
Ref. \cite{garciabellido}. This Lagrangian ($48$) introduces additional
scalar degrees of freedom by coupling the scalar field to the metric, which
is possible due to the appropriate behavior of the scalar field under
Lorentz transformations. This procedure gave rise to scalar-tensor theories
that correspond to generalizations of Jordan-Brans-Dicke theories and to
theories that lead to second-order field equations known as Horndeski
theories \cite{horn}, \cite{galileo}.

Finally it might be interesting to note that this same compactification
procedure applied to the AdS-Chern-Simons action \cite{cham1}-\cite{zan2} in
five dimensions leads to the Einstein-Hilbert-Cartan action in four
dimensions \cite{lepe} (see appendix $B$).

This work was supported in part by\textit{\ }FONDECYT Grants\textit{\ }No.%
\textit{\ }$1180681$ and No.\textit{\ }$1211219$ from the Government of
Chile. V.O and DS were supported in part by Concepci\'{o}n University, Chile.

\section{Appendix A: Randall-Sundrum Procedure}

In order to obtain an action for a $4$-dimensional gravity theory from the
Einstein-Chern-Simons action we will consider the following $5$-dimensional
Randall Sundrum type metric \cite{rand1,rand2,rd,gomez} 
\begin{eqnarray}
ds^{2} &=&e^{2f(\phi )}\tilde{g}_{\mu \nu }(\tilde{x})d\tilde{x}^{\mu }d%
\tilde{x}^{\nu }+r_{c}^{2}d\phi ^{2}  \notag \\
&=&e^{2f(\phi )}\tilde{\eta}_{mn}\tilde{e}^{m}\tilde{e}^{n}+r_{c}^{2}d\phi
^{2},
\end{eqnarray}%
where $e^{2f(\phi )}$ is the so-called "warp factor", and $r_{c}$ is the
so-called "compactification radius" of the extra dimension, which is
associated with the coordinate $0\leqslant \phi <2\pi $. The symbol $\sim $
denotes $4$-dimensional quantities related to the space-time $\Sigma _{4}.$
We will use the usual notation, 
\begin{eqnarray}
x^{\alpha } &=&\left( \tilde{x}^{\mu },\phi \right) ;\text{ \ \ \ \ }\alpha
,\beta =0,...,4;\text{ \ \ \ \ }a,b=0,...,4;  \notag \\
\mu ,\nu &=&0,...,3;\text{ \ \ \ \ }m,n=0,...,3;  \notag \\
\eta _{ab} &=&diag(-1,1,1,1,1);\text{ \ \ \ \ }\tilde{\eta}%
_{mn}=diag(-1,1,1,1).
\end{eqnarray}

This allows us, for example, to write 
\begin{eqnarray}
e^{m}(\phi ,\tilde{x}) &=&e^{f(\phi )}\tilde{e}^{m}(\tilde{x})=e^{f(\phi )}%
\tilde{e}_{\text{ }\mu }^{m}(\tilde{x})d\tilde{x}^{\mu };\text{ \ \ }%
e^{4}(\phi )=r_{c}d\phi .  \notag \\
k^{mn}(\phi ,\tilde{x}) &=&\tilde{k}^{mn}(\tilde{x})\text{, }k^{m4}=k^{4m}=0%
\text{,}  \label{s2}
\end{eqnarray}%
where, following Randall and Sundrum \cite{rand1,rand2} matter fields are
null in the fifth dimension.

From the vanishing torsion condition%
\begin{equation}
T^{a}=de^{a}+\omega _{\text{ }b}^{a}e^{b}=0,  \label{2t}
\end{equation}%
we obtain 
\begin{equation}
\omega _{\text{ }b\alpha }^{a}=-e_{\text{ }b}^{\beta }\left( \partial
_{\alpha }e_{\text{ }\beta }^{a}-\Gamma _{\text{ }\alpha \beta }^{\gamma }e_{%
\text{ }\gamma }^{a}\right) ,  \label{3t}
\end{equation}%
where $\Gamma _{\text{ }\alpha \beta }^{\gamma }$ is the Christoffel symbol.

From equations (\ref{s2}) and (\ref{2t}), we find%
\begin{equation}
\omega _{\text{ }4}^{m}=\frac{e^{f}f^{\prime }}{r_{c}}\tilde{e}^{m},\text{
with }f^{\prime }=\frac{\partial f}{\partial \phi },  \label{102t}
\end{equation}%
and the $4$-dimensional vanishing torsion condition 
\begin{equation}
\tilde{T}^{m}=\tilde{d}\tilde{e}^{m}+\tilde{\omega}_{\text{ }n}^{m}\tilde{e}%
^{n}=0,\text{ with \ }\tilde{\omega}_{\text{ }n}^{m}=\omega _{\text{ }n}^{m}%
\text{ \ and }\tilde{d}=d\tilde{x}^{\mu }\frac{\partial }{\partial \tilde{x}%
^{\mu }}.  \label{1030t}
\end{equation}

From (\ref{102t}), (\ref{1030t}) and the Cartan's second structural
equation, $R^{ab}=d\omega ^{ab}+\omega _{\text{ }c}^{a}\omega ^{cb}$, we
obtain the components of the $2$-form curvature%
\begin{equation}
R^{m4}=\frac{e^{f}}{r_{c}}\left( f^{\prime 2}-f^{\prime \prime }\right)
d\phi \tilde{e}^{m},\text{ \ }R^{mn}=\tilde{R}^{mn}-\left( \frac{%
e^{f}f^{\prime }}{r_{c}}\right) ^{2}\tilde{e}^{m}\tilde{e}^{n},\text{\ }
\label{105t}
\end{equation}%
where the $4$-dimensional $2$-form curvature is given by%
\begin{equation}
\tilde{R}^{mn}=\tilde{d}\tilde{\omega}^{mn}+\tilde{\omega}_{\text{ }p}^{m}%
\tilde{\omega}^{pn}.
\end{equation}

\section{Appendix B: Gravity in $4D$ from AdS-Chern-Simons gravity}

The AdS-Chern-Simons gravity \cite{cham1}-\cite{zan2} is a gauge theory
whose Lagrangian density is given by a 5-dimensional Chern-Simons form for
the Anti-de Sitter algebra, 
\begin{equation}
S_{EGB}=\frac{l^{2}}{8\kappa _{5}}\int \varepsilon _{abcde}\left(
R^{ab}R^{cd}e^{e}+\frac{2}{3l^{2}}R^{ab}e^{c}e^{d}e^{e}+\frac{1}{5l^{4}}%
\,e^{a}e^{b}e^{c}e^{d}e^{e}\right) ,  \label{ap3}
\end{equation}%
where $R^{ab}=\mathrm{d}\omega ^{ab}+\omega _{\text{ }c}^{a}\omega ^{cb}$
and $T^{a}=\mathrm{d}e^{a}+\omega _{\text{ }c}^{a}e^{c}$. The Lagrangian of
this action contains the Gauss-Bonnet term $L_{GB}$, the Einstein-Hilbert
term $L_{EH}$ and a cosmological term $L_{\Lambda }$. Using (\ref{s2}) and (%
\ref{105t}) we find

\begin{eqnarray}
L_{GB} &=&\varepsilon _{abcde}R^{ab}R^{cd}e^{e},  \notag \\
&=&r_{c}d\phi \left\{ \tilde{\varepsilon}_{mnpq}\tilde{R}^{mn}\tilde{R}%
^{pq}-\left( \frac{2e^{2f}}{r_{c}^{2}}\right) \left( 3f^{\prime
2}+2f^{\prime \prime }\right) \tilde{\varepsilon}_{mnpq}\tilde{R}^{mn}\tilde{%
e}^{p}\tilde{e}^{q}\right.  \notag \\
&&\text{ \ \ \ \ \ \ \ \ \ \ \ \ \ \ }\left. +\left( \frac{e^{4f}}{r_{c}^{4}}%
f^{\prime 2}\right) \left( 5f^{\prime 2}+4f^{\prime \prime }\right) \tilde{%
\varepsilon}_{mnpq}\tilde{e}^{m}\tilde{e}^{n}\tilde{e}^{p}\tilde{e}%
^{q}\right\} ,  \label{4t'}
\end{eqnarray}

\begin{eqnarray}
L_{EH} &=&\varepsilon _{abcde}R^{ab}e^{c}e^{d}e^{e},  \notag \\
&=&r_{c}d\phi \left\{ \left( 3e^{2f}\right) \tilde{\varepsilon}_{mnpq}\tilde{%
R}^{mn}\tilde{e}^{p}\tilde{e}^{q}\right.  \notag \\
&&-\left. \left( \frac{e^{4f}}{r_{c}^{2}}\right) \left( 5f^{\prime
2}+2f^{\prime \prime }\right) \tilde{\varepsilon}_{mnpq}\tilde{e}^{m}\tilde{e%
}^{n}\tilde{e}^{p}\tilde{e}^{q}\right\} ,  \label{4t''}
\end{eqnarray}

\begin{eqnarray}
L_{\Lambda } &=&\varepsilon _{abcde}e^{a}e^{b}e^{c}e^{d}e^{e},  \notag \\
&=&5r_{c}d\phi e^{4f}\tilde{\varepsilon}_{mnpq}\tilde{e}^{m}\tilde{e}^{n}%
\tilde{e}^{p}\tilde{e}^{q}.  \label{4t'''}
\end{eqnarray}

Replacing (\ref{4t'}), (\ref{4t''}) and (\ref{4t'''}) in (\ref{ap3}) we
obtain 
\begin{equation}
\tilde{S}[\tilde{e}]=\frac{1}{8\kappa _{5}}\int_{\Sigma _{4}}\tilde{%
\varepsilon}_{mnpq}\left( \tilde{A}\tilde{R}^{mn}\tilde{e}^{p}\tilde{e}^{q}+%
\tilde{B}\ \tilde{e}^{m}\tilde{e}^{n}\tilde{e}^{p}\tilde{e}^{q}\right) ,
\label{999}
\end{equation}%
where%
\begin{eqnarray}
\tilde{A} &=&\frac{2\pi l^{2}}{r_{c}}\left( 1+\frac{r_{c}^{2}}{l^{2}}\right)
,  \label{20t} \\
\tilde{B} &=&-\frac{\pi }{4r_{c}}\left( \frac{l^{2}}{r_{c}^{2}}-2-3\frac{%
r_{c}^{2}}{l^{2}}\right) .  \label{21t}
\end{eqnarray}%
It is direct to see that the action (\ref{999}) lead to the
Einstein-Hilbert-Cartan action when 
\begin{equation}
\tilde{A}=6\pi ^{2}r_{c}\text{ \ \ }and\text{ \ \ }\tilde{B}=-\pi
^{2}\Lambda _{4D}r_{c}.  \label{21t'}
\end{equation}%
From (\ref{20t}), (\ref{21t}) and (\ref{21t'}) we have

\begin{equation}
\frac{r_{c}^{2}}{l^{2}}=\frac{1}{3\pi -1},  \label{24t'}
\end{equation}%
and then \ 
\begin{equation}
\text{ }\Lambda _{4D}=\Lambda _{4D}\left( r_{c}\right) =\left( \frac{3\pi -4%
}{3\pi -1}\right) \frac{3}{4r_{c}^{2}}.  \label{25tc}
\end{equation}

Introducing (\ref{21t'}) in (\ref{999}) we obtain

\begin{equation}
\tilde{S}[\tilde{g}]=\int d^{4}\tilde{x}\sqrt{-\tilde{g}}\left( \tilde{R}%
-2\Lambda _{4D}\right) ,  \label{31t'}
\end{equation}%
whose field equations are 
\begin{equation}
G_{\mu \nu }=-\Lambda _{4D}g_{\mu \nu },
\end{equation}%
where we have used (\ref{1}, \ref{2}).

\section{Appendix C: The most general scalar-tensor theories of gravity}

In reference \cite{garciabellido} the most general scalar-tensor theory was
built. This theory satisfies the following two conditions $(i)$ it is
described by a principle of action in which the Lagrange function is a $D$%
-form invariant under local Lorentz transformations on a pseudo-Riemannian
manifold $M$; $\left( ii\right) $ the Lagrange function is constructed from
the exterior product of the $1$-form vielbein $e^{a}$, the $2$-form
curvature $R^{ab}$, the first derivatives of the scalar field $\Psi ^{a}$,
and the second derivatives of the scalar field $\Phi ^{a}$, where the fields 
$\Phi ^{a}$ and $\Psi ^{a}$ are defined in terms of the $0$-form $\varphi $,
as $\Phi ^{a}\equiv \nabla ^{a}\nabla _{b}\phi e^{b}$, $\Psi ^{a}\equiv
\nabla ^{a}\phi \nabla _{b}\phi e^{b}$.

The action for said theory is given by 
\begin{equation}
S=\sum_{l,m,n}^{p\leq D}\int_{M}\alpha _{lmn}\mathcal{L}_{(lmn)},  \label{a1}
\end{equation}%
where $p=2l+m+n$, con $n\leq 1,$ $\alpha _{lmn}=\alpha _{lmn}\left( \phi ,X,%
\left[ \Phi \right] ,\cdot \cdot \cdot \right) $ are $0$-forms, with $%
X=\left( -1/2\right) \phi ^{,\mu }\phi _{,\mu }$, $\left[ \Phi \right] =\phi
_{\text{ };a}^{;a}$, and 
\begin{equation}
\mathcal{L}_{(lmn)}=\theta _{a_{1}b_{1}\cdot \cdot \cdot
a_{l}b_{l}c_{1}\cdot \cdot \cdot c_{m}d_{1}\cdot \cdot \cdot d_{n}}^{\ast
}\dbigwedge\limits_{i=1}^{l}R^{a_{i}b_{i}}\wedge
\dbigwedge\limits_{j=1}^{m}\Phi ^{c_{j}}\wedge
\dbigwedge\limits_{k=1}^{n}\Psi ^{d_{k}},  \label{a2}
\end{equation}%
where $\tbigwedge $ is a shorthand for a consecutive set of wedge products, $%
l,m,n\in 
\mathbb{N}
,$ and $\theta _{a_{1}\cdot \cdot \cdot a_{k}}^{\ast }$ is defined as 
\begin{equation}
\theta _{a_{1}\cdot \cdot \cdot a_{k}}^{\ast }=\frac{1}{(D-k)!}\varepsilon
_{a_{1}\cdot \cdot \cdot a_{k}a_{k+1}\cdot \cdot \cdot
a_{D}}e^{a_{k+1}}\wedge \cdot \cdot \cdot \wedge e^{a_{D}}.  \label{a3}
\end{equation}

For the case $p=4$ with $n=0$, it was found that (see equation $48$ of \cite%
{garciabellido})

\begin{eqnarray}
\mathcal{L}_{6}^{NH} &=&E_{6}\left(
R_{abcd}R^{abcd}-4R_{ab}R^{ab}+R^{2}\right) \eta   \notag \\
&&+2E_{6,X}\left\{ \left( \left[ \Phi \right] ^{2}-\left[ \Phi ^{2}\right]
\right) R+4\left( \left[ \Phi \right] \Phi _{ab}-\Phi _{ab}^{2}\right)
R^{ab}\right.   \notag \\
&&+\left. 2\Phi _{ab}\Phi _{cd}R^{abcd}\right\} \eta +\frac{1}{3}%
E_{6,XX}\left( \left[ \Phi \right] ^{4}-6\left[ \Phi \right] ^{2}\left[ \Phi
^{2}\right] +3\left[ \Phi ^{2}\right] ^{2}\right.   \notag \\
&&\left. +8\left[ \Phi \right] \left[ \Phi ^{3}\right] -6\left[ \Phi ^{4}%
\right] \right) \eta ,  \label{a4}
\end{eqnarray}%
where $\left[ \Phi \right] =\phi _{\text{ \ };a}^{;a}$, $\Phi _{ab}=\phi
_{;ab}$, $\Phi _{ab}^{n}=\phi _{;ac_{1}}\phi _{\text{ \ };c_{2}}^{;c_{1}}%
\cdot \cdot \cdot \phi _{\text{ \ \ \ \ \ };b}^{;c_{n-1}}$, so that 
\begin{eqnarray}
\left[ \Phi \right] ^{2}-\left[ \Phi ^{2}\right]  &=&\phi _{\text{ \ }%
;a}^{;a}\phi _{\text{ \ };b}^{;b}-\phi _{\text{ \ };b}^{;a}\phi _{\text{ \ }%
;a}^{;b}  \notag \\
\left[ \Phi \right] \Phi _{ab}-\Phi _{ab}^{2} &=&\phi _{\text{ \ }%
;d}^{;d}\phi _{;ab}-\phi _{;ac}\phi _{\text{ \ };b}^{;c}  \notag \\
\Phi _{ab}\Phi _{cd} &=&\phi _{;ab}\phi _{;cd}
\end{eqnarray}%
\begin{eqnarray}
&&\left[ \Phi \right] ^{4}-6\left[ \Phi \right] ^{2}\left[ \Phi ^{2}\right]
+3\left[ \Phi ^{2}\right] ^{2}+8\left[ \Phi \right] \left[ \Phi ^{3}\right]
-6\left[ \Phi ^{4}\right]   \notag \\
&=&\phi _{\text{ \ };a}^{;a}\phi _{\text{ \ };b}^{;b}\phi _{\text{ \ }%
;c}^{;c}\phi _{\text{ \ };d}^{;d}-6\phi _{\text{ \ };a}^{;a}\phi _{\text{ \ }%
;b}^{;b}\phi _{\text{ \ };d}^{;c}\phi _{\text{ \ };c}^{;d}+3\phi _{\text{ \ }%
;b}^{;a}\phi _{\text{ \ };a}^{;b}\phi _{\text{ \ };d}^{;c}\phi _{\text{ \ }%
;c}^{;d}  \notag \\
&&+8\phi _{\text{ \ };a}^{;a}\phi _{\text{ \ };c}^{;b}\phi _{\text{ \ }%
;d}^{;c}\phi _{\text{ \ };b}^{;d}-6\phi _{\text{ \ };b}^{;a}\phi _{\text{ \ }%
;c}^{;b}\phi _{\text{ \ };d}^{;c}\phi _{\text{ \ };a}^{;d}.
\end{eqnarray}

The corresponding equations of motion do not have higher derivatives. This
theory, whose Lagrangian is $\mathcal{L}_{6}^{NH}=\mathcal{L}_{6}^{NH}\left[
E_{6}\left( \phi ,X\right) \right] $, was called by its authors \cite%
{garciabellido} "kinetic Gauss-Bonnet theory". In the particular case when $%
E_{6}=E_{6}\left( \phi \right) $ the theory corresponds to Gauss-Bonnet
gravity coupled to a scalar function \cite{sn}. This result allowed the
authors of Ref. \cite{garciabellido} to prove that said theory is contained
in Horndeski's theory \cite{horn}, and that when there is a dependence only
on the kinetic term on the coefficient $E_{6}=E_{6}\left( X\right) $, it
follows that the kinetic Gauss-Bonnet Lagrangian is an exact form.

\end{document}